\documentclass[aps,prl,twocolumn,groupedaddress]{revtex4} 

\usepackage{graphicx}
\usepackage{bm}

\begin{document}


\title{Photoassociative spectroscopy at long range in
ultracold strontium}


\author{S. B. Nagel, P. G. Mickelson, A. D. Saenz, Y. N. Martinez,  Y. C. Chen, and T. C. Killian}

\affiliation{Rice University, Department of Physics and Astronomy
and Rice Quantum Institute, Houston, Texas, 77251}

\author{P. Pellegrini, and R. C\^ot\'e}

\affiliation{Department of Physics, U-3046, University of Connecticut, Storrs, CT, 06269-3046}


\date{\today}

\begin{abstract}
We report photoassociative spectroscopy of $^{88}$Sr$_2$ in a magneto-optical
trap operating on the ${^1S_0}\rightarrow{^3P_1}$ intercombination
line at 689\,nm. Photoassociative transitions are driven with
a laser red-detuned by 600-2400\,MHz
from the ${^1S_0}\rightarrow{^1P_1}$ atomic resonance at 461\,nm.
Photoassociation takes place at
extremely large internuclear separation, and the photoassociative spectrum is
strongly affected by relativistic retardation.
A fit of the transition frequencies determines the
${^1P_1}$ atomic lifetime ($\tau=5.22 \pm 0.03$\,ns) and resolves a discrepancy between experiment and
recent theoretical calculations.

\end{abstract}


\maketitle




    Photoassociative spectroscopy (PAS) in laser-cooled
    gases \cite{ljp96}
    is a powerful probe of molecular potentials and
    atomic cold collisions.
   It provides
    accurate determination of ground state scattering lengths
    and  excited state lifetimes \cite{wbz99}.
    Photoassociation occurs naturally in laser cooling and trapping
    experiments in which
    the lasers are red-detuned from atomic resonance, so characterizing
   the process is also important for understanding and optimizing the
   production of ultracold atoms (e.g. \cite{dva99}).



    PAS of alkaline earth atoms differs
    significantly from more common studies
    of alkali metal atoms.
    The most abundant isotopes of
    alkaline earth atoms lack hyperfine structure,
    making these systems ideal for
    testing PAS theory. Recent theoretical
    work \cite{mjs01} emphasized the
    ability to resolve transitions at extremely large internuclear
    separation and very small detuning from the atomic asymptote.
    The finite speed of light
    modifies the potential in this regime through %
    relativistic
    retardation \cite{mea68,jjl96,mah96}.

    There is also great interest in alkaline-earth-atom cold collisions because
    of their importance for optical frequency
    standards \cite{wbd02,coh03,ika03,tka03}
     and for the creation
    of quantum degenerate gases \cite{dpk02,iik00,tmk03}.
   In addition, collisions involving metastable states
   \cite{ctk04,der01,ksg03,sgr03} display
    novel properties that arise from
    electric quadrupole-quadrupole  or magnetic dipole-dipole
   interactions.

      PAS red-detuned from  the principle transitions in calcium  \cite{dbw03} and
      ytterbium \cite{tkh04} is well
      characterized, resulting in accurate measurements of the
     first excited $^1P_1$ lifetimes and  the ground
     state $s$-wave scattering lengths.
     In spite of its importance for potential optical frequency standards,
      little is known about strontium.
     The photoassociative loss rate induced
     by trap lasers in a ${^1S_0}\rightarrow{^1P_1}$
     magneto-optical trap (MOT)
     has
     been measured \cite{dva99}, and preliminary results from more extensive
     spectroscopy were recently reported  \cite{ykt04}.
     {\it Ab initio} strontium potentials
     have been calculated for small internuclear
     separation ($R<9$\,nm) \cite{baa96}.

\begin{figure}
  \includegraphics[width=3in,clip=true]{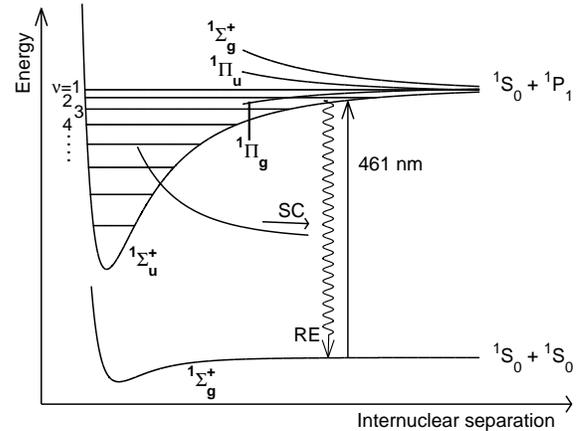}\\
  \caption{Molecular levels and photoassociative transitions.
  Resonant excitation occurs near the outer molecular turning point
  of states of the $^1\Sigma_u$ potential, and
  it can lead to
  trap loss through two processes.
  The molecule can radiatively
  decay at smaller internuclear separation to two ground  state atoms
  with kinetic energies exceeding the trap depth. This is known as
  radiative escape (RE). In a state-changing collision (SC), at small internuclear separation
  the molecular state changes to one corresponding  at long range to a
  lower-lying electronic configuration of free atoms. The atoms
  exit the collision with greatly increased kinetic energy and escape the trap.
  }\label{PASdrawing}
\end{figure}

 In this paper we report PAS of $^{88}$Sr near
 the ${^1S_0}\rightarrow{^1P_1}$ atomic resonance at 461\,nm
 (Fig. \ref{PASdrawing}). The simple spectrum allows us to resolve
 transitions as little as 600\,MHz detuned from the atomic resonance, which produces
 molecules with greater internuclear separation than in any previous PAS work.
 Our determination of the  first excited $^1P_1$ lifetime resolves a discrepancy
 between experiment and recent theoretical work, and provides an importance test
 of atomic structure theory for alkaline earth atoms.

    The most common form of PAS involves resonantly inducing trap loss
    in a MOT, although early work was also conducted
     in optical-dipole traps \cite{ljp96,wbz99}. Recent experiments have also
    studied spectroscopy of atoms in magnetic traps, especially in Bose-Einstein
    condensates \cite{ppj03,mdh02}. The experiments described in this article
    were performed in a MOT, but the MOT
operated on the ${^1S_0}\rightarrow{^3P_1}$ intercombination
line at 689\,nm  \cite{kii99}, rather
    than an electric-dipole allowed transition.
    This results in lower atom temperature, a shallower trap,
    and higher atom density
    than a standard MOT.


Atoms are initially trapped in a MOT operating on the
461\,nm  ${^1S_0}\rightarrow{^1P_1}$ transition, as described in
 \cite{nsl03}.
During the
loading phase,
the peak intensity in each MOT beam is
$6$ mW$/$cm$^2$, 
and the axial magnetic field gradient generated by a pair of
anti-Helmholtz coils is $56$ G$/$cm.
 The intensity is then reduced by about a factor of 8 for  $3.5$\,ms
 to reduce the atom temperature.
 After this stage the MOT contains about
 $150$ million atoms at  $2$\,mK.

The 461\,nm laser cooling light is then extinguished, the
field gradient is reduced to 
$2.1$ G/cm, and the $689$ nm light for the $^1$S$_0$ $\rightarrow$ $^3$P$_1$
intercombination-line MOT is switched on.
This MOT also consists of three retro-reflected beams, each with
a diameter of 2\,cm and intensity of $400$\,$\mu$W/cm$^2$.
The frequency of the 689\,nm laser is detuned
from the atomic resonance by 0.5 MHz and spectrally broadened with a
$\pm$400\,kHz
sine-wave modulation. Transfer and equilibration take about 50\,ms, after which
there are 15 million atoms at a temperature of 5\,$\mu$K,
peak density of about $5\times 10^{11}$\,cm$^{-3}$,
and $1/\sqrt{{\rm e}}$ density radius of about 100\,$\mu$m.

The intercombination-line MOT operates
for  an adjustable hold time 
before measuring the remaining number of atoms
with absorption imaging using the $^1$S$_0$ $\rightarrow$ $^1$P$_1$ transition.
The lifetime of atoms in the MOT is approximately 350\,ms, limited
by background gas collisions. To detect photoassociative resonances,
a PAS laser is applied to the atoms during
hold times of $300-400$\,ms. When the PAS laser is tuned to a molecular
resonance, photoassociation provides another loss mechanism for the MOT,
decreasing the number of atoms.


Light for photoassociation  is derived from the same laser that
produces the 461\,nm light for laser cooling.
Several acousto-optic modulators (AOM),
detune the light 600 to 2400\,MHz
to the red of the atomic transition.  The laser frequency is locked relative
to a Doppler-free saturated-absorption feature in a vapor cell, with an
uncertainty of about  2 MHz.
 The last
AOM in the offset chain,
in a double-pass configuration,  controls the power
of the  PAS
laser and scans the frequency up to 200\, MHz with minimal beam misalignment.
PAS light is double-passed through the MOT in a standing wave, with a
1/e$^2$ intensity radius of $w=3$\,mm.

\begin{figure}
  \includegraphics[width=3.35in]{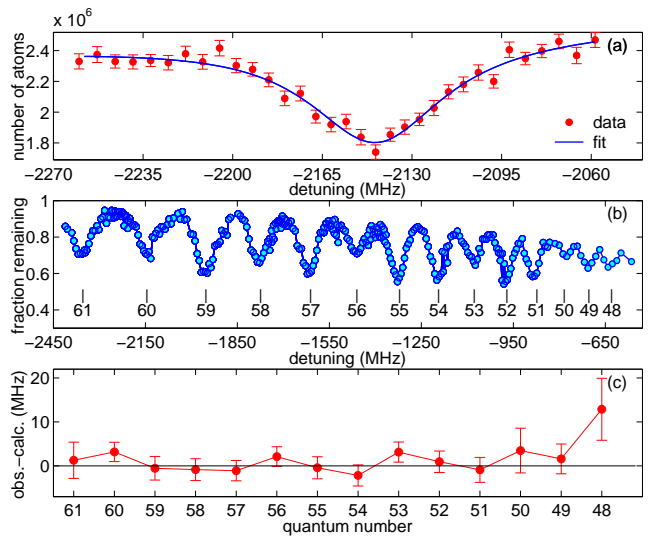}\\
  \caption{(a) Spectrum for photoassociation
  to the $v=60$ molecular state. Detuning is
  with respect to the atomic ${^1S_0}-{^1P_1}$ resonance.
 The PAS laser intensity is $1.4$\,mW/cm$^2$.
 The Lorentzian
 fit yields a FWHM linewidth of $70\pm 7$\,MHz.
 (b) Representative spectra for all
  transitions observed in this study.
  Lines connect the data points to guide the eye.
  Conditions
vary for the individual scans that make up the full spectrum,
so amplitudes
and linewidths should not be quantitatively compared. The baseline has been
adjusted to match the expected curve for overlapping Lorentzians.
(c) Differences between  experimental and calculated positions of the 14 measured levels.
  }\label{fullspectrum}
\end{figure}

Figure \ref{fullspectrum}A shows
the number of atoms remaining after
the hold time 
as a function of
PAS laser detuning near a typical resonance.
 We record
spectra with PAS laser intensity from $0.5-5$\,mW/cm$^2$, much less than the
atomic saturation intensity ($I_{sat}=40$\,mW/cm$^2$). For the
lower intensities, the linewidth is close to
the expected natural linewidth
of twice
the inverse of the ${^1P_1}$ atomic lifetime
($\tau^{-1}=  2\pi \times 30.5$\,MHz).
Thermal broadening ($k_B T/h\approx 100$\,kHz) is negligible.
Only $s$-wave collisions occur, so only a single  rotational level
($J=1$) is excited.
For higher intensity, the observed line is broadened slightly
because, on resonance, the signal saturates if
a large fraction of atoms is lost.


Figure \ref{fullspectrum}B shows a complete spectrum of  the
$14$ different PAS resonances observed in this study.
Center frequencies for each transition are determined  by
Lorentzian fits.
The typical statistical uncertainty is 2-3\,MHz, but
there is approximately 2\,MHz additional uncertainty  arising from the
lock of the laser.
The Condon
radius for the excitation varies from
20\,nm for the largest  detuning to
32\,nm for the smallest. These extremely large values
exceed the inter-nuclear spacing
of molecules formed in photoassociative spectroscopy to
pure long-range potentials \cite{jjl96,fcc98}.

The region of the attractive $^1\Sigma^+_u$ molecular potential probed
by PAS corresponds to large internuclear separations,
and is typically approximated
by
\begin{equation}\label{potential}
    V(R) = D-\frac{C_3}{R^3} + \frac{\hbar^2 [J(J+1)+2]}{2\mu R^2}, \;
    C_3=\frac{3 \hbar \lambda^3}{16 \pi^3 \tau},
    \label{eq:V(R)}
\end{equation}
where $D$ is the dissociation energy, $\mu$ the reduced mass, and
$\lambda = 461$ nm.
However, at  very large separations, the
atom-atom interaction is modified by relativistic corrections.
This retardation effect is well understood \cite{mea68}
and can be included in the analysis of the spectrum through
$C_3\rightarrow C_3 [\cos(u)+u\sin(u)]$, where $u =2\pi R / \lambda$.
Machholm {\it et al.} \cite{mjs01} discussed this in the context of
PAS of alkaline earth atoms.

To extract molecular parameters from the positions of the PAS resonances,
we constructed a
potential curve by smoothly connecting the long range form (Eq.\ (\ref{eq:V(R)}))
to a short range curve at a distance of 1.5\,nm. The short
range {\it ab initio} potential was obtained using a semi-empirical
two-valence-electron pseudopotential method \cite{baa96}.
To account for uncertainty
in the short range potential, the
position of the repulsive wall was varied as a fit parameter.
 The
rovibrational levels in the $^1\Sigma_u^+$ potential were found by solving
the radial Schr\"odinger equation using the Mapped Fourier
Grid Method \cite{kdk99} with a grid size typically larger than 500\,nm
containing about 1000 grid points.

\begin{figure}
  \includegraphics[width=1.75in, angle=270]{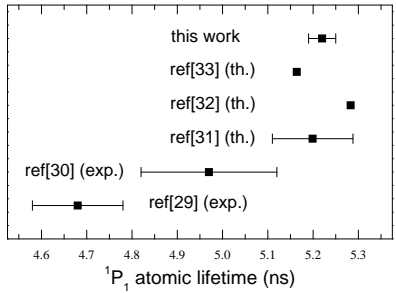}\\
   \caption{Comparison of experimental (exp.) and theoretical (th.)
  values for the lifetime of the $^1P_1$
  level.}\label{lifetime}
\end{figure}

We found that
the observed levels range from 48 to 61, starting the
numbering from the dissociation limit. The best
fit of the data  was achieved with a value of $C_3=18.54$\,a.u., with a
reduced chi-squared value of
$\chi^2=0.79$.
(See
Fig.\ \ref{fullspectrum}C). $\chi^2$ varied by less than 10\%
for a change of level assignments of $\Delta v = \pm 1$, so we
consider our assignment to be uncertain by this amount. The value
of $C_3$ changed by $\pm 0.5$\% as the assignments changed, which is
much larger  than the statistical uncertainty for a given
assignment. We thus take $\pm 0.5$\% as our uncertainty in $C_3$.
A fit to the level spacings, as opposed to the absolute positions, yielded
the same value of $C_3$. From $C_3$, we derive a natural decay rate of
the atomic $5p$ $^1P_1$ state of $\tau=5.22 \pm 0.03$\,ns.
Decay channels other than $^1P_1\rightarrow {^1S_0}$ can be neglected
at this level of accuracy. The most recent experimental determinations of
$\tau$ use the Hanle effect  \cite{kma80,ldg64}.
Our result agrees well with recent theoretical
values \cite{pkr01,sjo02,mbr03} (Fig.\ \ref{lifetime}).

Retardation effects shift the levels by approximately
100 MHz in this regime, which is similar to the shift seen in a pure long
range potential in Na$_2$ \cite{jjl96}. If retardation effects are neglected and the
data is fit using the simple LeRoy
semiclassical treatment \cite{ler80},
the level assignments change significantly, and $C_3$ changes by more than 10\%,
putting it
outside the range of recent theoretical results.
 Center frequency shifts due to
coupling to the continuum of the ground state have been ignored
as they are less than 50 kHz for typical experimental
temperatures.



 \begin{figure}
   \includegraphics[width=3.35in]{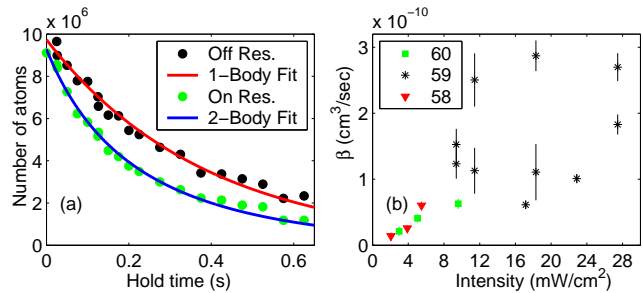}\\
  \caption{(a) Number of atoms as a function of time for atoms in the
  intercombination-line MOT. The PAS laser is tuned near the
  $v=59$ molecular resonance. 
  In one data set the PAS laser
  is on resonance, 
  and in the other
  it is tuned to the blue of resonance
  by 42\,MHz,
  and loss due to photoassociation is small.
  (b) $\beta$ as a function of PAS laser intensity for several
  molecular resonances. Error
  bars denote statistical uncertainties.
  }\label{beta}
\end{figure}

Figure\ \ref{beta}A shows an
example of the number of atoms, $N$, in the intercombination-line MOT
as a function of time.
The density in the MOT varies according to
$\dot{n}=-\Gamma n-\beta n^2$.
Integrating this equation over volume and solving the
differential equation
yields
\begin{equation}\label{number}
   N(t)={N_0 \rm{e}^{-\Gamma t} \over 1+{N_0 \beta \over \Gamma (2\sqrt{2\pi}\sigma)^3}(1-\rm{e}^{-\Gamma t})},
\end{equation}
where  we approximate the density as
$n(r)=n_0 \rm{e}^{-r^2/2\sigma^2}$, and $\Gamma$ is the one-body loss
due primarily to background gas collisions.
$\beta$ is the two-body loss rate
and it contains important
information about the dynamics of photoassociation and trap loss.



Even with photoassociation, the deviation of $N(t)$ from a
single exponential decay is  small,
making it difficult to independently extract $\beta$ and $\Gamma$
 with high accuracy from a single decay curve.
 To address this problem, we take advantage of the
 fact that when the PAS laser is not on a molecular resonance,
  the photoassociation rate is small (See Fig.\ \ref{fullspectrum}B),
  and within the accuracy of our
  measurement it can be neglected.
  All other processes, however, such as off-resonance scattering from the
  atomic transition, are the same. We thus take data in pairs
  of on and off-resonance, and fit off-resonance data
  to $N(t)=N_0 \rm{e}^{-\Gamma t}$. The on-resonance data
  is then fit with Eq.\ (\ref{number}) with
  $\Gamma$ fixed to the value determined from the off-resonance
data. The resulting two body decay rates are shown in Fig.\ \ref{beta}B.
For the relatively small intensities used, $\beta$ is expected to vary linearly with
laser power \cite{sbt98}.

We
 currently
image along the direction of gravity
and lack the additional diagnostic required to obtain information
on the third dimension of the atom cloud.
Gravity can  distort the equilibrium shape of
the intercombination-line MOT  because the light force
is so weak. However,  we operate the MOT in the regime
where the laser detuning is comparable to the modulation of the
laser spectrum, and, as shown in  \cite{lil04}, in this regime
the effect is small.
Based on the range of distortions we observe in the imaged dimensions
for various
misalignments of the MOT beams,
we  allow for a scale uncertainty of a factor of two
in the volume of the MOT.
This contributes an identical uncertainty in the measurements of
$\beta$.

Measurements of MOT size suggest that as
 the PAS laser power increases,
 the cloud radius increases slightly (20\% for the highest
 intensities),  decreasing the atomic density.
This is expected because
off-resonance scattering from the
atomic transition should heat the sample.
The data quality does not allow a reliable correction for this
effect, however, and we use the average of all observed $\sigma$
as a constant $\sigma$ in Eq.\ \ref{number}. This  contributes another
approximately 50\% uncertainty in our determination of $\beta$. Overall,
we quote a factor of 3 uncertainty due to systematic effects, which
dominates over statistical error.

The measured values of $\beta \approx 1 \times 10^{-11}$\,cm$^3$/sec for
1\,mW/cm$^2$ in the intercombination-line MOT
are comparable to
theoretical \cite{mjs01}
and experimental \cite{dva99} values
of $\beta$ for  atoms
in a MOT operating on the 461\,nm transition. This provides some insight
into the typical kinetic energy of the resulting atoms after radiative decay
of a molecule to the free-atom continuum.

A MOT based on  an electric-dipole
allowed transition would typically have a trap depth
of about $1$\,K. In our experiment, the transfer efficiency
of atoms from the
461\,nm MOT  places a lower limit of $0.5$\,mK on the
depth of the intercombination-line MOT.
Using the formalism of \cite{lil04} and the parameters of the intercombination-line MOT,
we calculate a trap depth of 2\,mK.
As described in Fig.\ \ref{PASdrawing},
photoassociative loss requires excitation to a molecular state
followed by decay to a configuration of two atoms with enough kinetic energy to
escape from the trap \cite{mjs01}. The fact that the observed linewidths
are close to the theoretical minimum of $1/(\pi\tau)$ suggests that radiative escape
dominates in this regime, as predicted in theoretical calculations of
PAS rates for magnesium \cite{mjs01}. If $\beta$ is similar for trap depths of $1$\,K and 1\,mK,
then the energy released during radiative decay must be on the order of
1\,K or greater. Extensive calculations of radiative escape in lithium \cite{cda98} also led to a
an estimate of a few kelvin for the energy released in this process.

Photoassociative spectroscopy  has elucidated dynamics of collisions
in an intercombination-line MOT and provided an accurate measurement
of the lifetime of the lowest $^1P_1$ state of strontium. This has resolved
a previous discrpenancy between experimental and theoretical values of the lifetime.
We have also taken advantage of the simple
structure of the spectrum to measure transitions at very large internuclear
separation where the
level spacing and natural linewidth are comparable and retardation effects
are large.

This research was supported by the Office for Naval Research, National
Science Foundation,
Welch Foundation,
Research Corporation,
Alfred P. Sloan Foundation, and David and Lucille Packard
Foundation.
The authors are grateful to
A. R. Allouche for providing them the $Sr_2$ ab initio potentials.


\begin{thebibliography}{36}
\expandafter\ifx\csname natexlab\endcsname\relax\def\natexlab#1{#1}\fi
\expandafter\ifx\csname bibnamefont\endcsname\relax
  \def\bibnamefont#1{#1}\fi
\expandafter\ifx\csname bibfnamefont\endcsname\relax
  \def\bibfnamefont#1{#1}\fi
\expandafter\ifx\csname citenamefont\endcsname\relax
  \def\citenamefont#1{#1}\fi
\expandafter\ifx\csname url\endcsname\relax
  \def\url#1{\texttt{#1}}\fi
\expandafter\ifx\csname urlprefix\endcsname\relax\def\urlprefix{URL }\fi
\providecommand{\bibinfo}[2]{#2}
\providecommand{\eprint}[2][]{\url{#2}}

\bibitem[{\citenamefont{Lett et~al.}(1996)\citenamefont{Lett, Julienne, and
  Phillips}}]{ljp96}
\bibinfo{author}{\bibfnamefont{P.~D.} \bibnamefont{Lett}},
  \bibinfo{author}{\bibfnamefont{P.~S.} \bibnamefont{Julienne}},
  \bibnamefont{and} \bibinfo{author}{\bibfnamefont{W.~D.}
  \bibnamefont{Phillips}}, \bibinfo{journal}{{ Annu. Rev. Phys. Chem.}}
  \textbf{\bibinfo{volume}{46}}, \bibinfo{pages}{423} (\bibinfo{year}{1996}).

\bibitem[{\citenamefont{Weiner et~al.}(1999)\citenamefont{Weiner, Bagnato,
  Zilio, and Julienne}}]{wbz99}
\bibinfo{author}{\bibfnamefont{J.}~\bibnamefont{Weiner}} \bibnamefont{{\it et al.}},
  \bibinfo{journal}{{ Rev. Mod. Phys.}} \textbf{\bibinfo{volume}{71}},
  \bibinfo{pages}{1} (\bibinfo{year}{1999}).

\bibitem[{\citenamefont{Dinneen et~al.}(1999)\citenamefont{Dinneen, Vogel,
  Arimondo, Hall, , and Gallagher}}]{dva99}
\bibinfo{author}{\bibfnamefont{T.~P.} \bibnamefont{Dinneen}} \bibnamefont{{\it et al.}},
  \bibinfo{journal}{{ Phys. Rev. A}} \textbf{\bibinfo{volume}{59}},
  \bibinfo{pages}{1216} (\bibinfo{year}{1999}).

\bibitem[{\citenamefont{Machholm et~al.}(2001)\citenamefont{Machholm, Julienne,
  and Suominen}}]{mjs01}
\bibinfo{author}{\bibfnamefont{M.}~\bibnamefont{Machholm}},
  \bibinfo{author}{\bibfnamefont{P.~S.} \bibnamefont{Julienne}},
  \bibnamefont{and} \bibinfo{author}{\bibfnamefont{K.-A.}
  \bibnamefont{Suominen}}, \bibinfo{journal}{{ Phys. Rev. A}}
  \textbf{\bibinfo{volume}{64}}, \bibinfo{pages}{033425}
  (\bibinfo{year}{2001}).

\bibitem[{\citenamefont{Meath}(1968)}]{mea68}
\bibinfo{author}{\bibfnamefont{W.~J.} \bibnamefont{Meath}}, \bibinfo{journal}{{
  J. Chem. Phys.}} \textbf{\bibinfo{volume}{48}}, \bibinfo{pages}{227}
  (\bibinfo{year}{1968}).

\bibitem[{\citenamefont{Jones et~al.}(1996)\citenamefont{Jones, Julienne, Lett,
  Phillips, Tiesinga, and Williams}}]{jjl96}
\bibinfo{author}{\bibfnamefont{K.~M.} \bibnamefont{Jones}} \bibnamefont{{\it et al.}},
  \bibinfo{journal}{{ Europhys. Lett.}} \textbf{\bibinfo{volume}{35}},
  \bibinfo{pages}{85} (\bibinfo{year}{1996}).

\bibitem[{\citenamefont{McAlexander et~al.}(1996)\citenamefont{McAlexander,
  Abraham, and Hulet}}]{mah96}
\bibinfo{author}{\bibfnamefont{W.~I.} \bibnamefont{McAlexander}},
  \bibinfo{author}{\bibfnamefont{E.~R.~I.} \bibnamefont{Abraham}},
  \bibnamefont{and} \bibinfo{author}{\bibfnamefont{R.~G.} \bibnamefont{Hulet}},
  \bibinfo{journal}{{ Phys. Rev. A}} \textbf{\bibinfo{volume}{54}},
  \bibinfo{pages}{5} (\bibinfo{year}{1996}).

\bibitem[{\citenamefont{Wilpers et~al.}(2002)\citenamefont{Wilpers, Binneweis,
  Degenhardt, Sterr, Helmcke, and Riehle}}]{wbd02}
\bibinfo{author}{\bibfnamefont{G.}~\bibnamefont{Wilpers}} \bibnamefont{{\it et al.}},
  \bibinfo{journal}{{Phys. Rev. Lett.}} \textbf{\bibinfo{volume}{89}},
  \bibinfo{pages}{230801} (\bibinfo{year}{2002}).

\bibitem[{\citenamefont{Curtis et~al.}(2003)\citenamefont{Curtis, Oates, and
  Hollberg}}]{coh03}
\bibinfo{author}{\bibfnamefont{E.~A.} \bibnamefont{Curtis}},
  \bibinfo{author}{\bibfnamefont{C.~W.} \bibnamefont{Oates}}, \bibnamefont{and}
  \bibinfo{author}{\bibfnamefont{L.}~\bibnamefont{Hollberg}},
  \bibinfo{journal}{{J. Opt. Soc. Am. B - Opt. Phys.}}
  \textbf{\bibinfo{volume}{20}}, \bibinfo{pages}{977} (\bibinfo{year}{2003}).

\bibitem[{\citenamefont{Ido and Katori}(2003)}]{ika03}
\bibinfo{author}{\bibfnamefont{T.}~\bibnamefont{Ido}} \bibnamefont{and}
  \bibinfo{author}{\bibfnamefont{H.}~\bibnamefont{Katori}}, \bibinfo{journal}{{
  Phys. Rev. Lett.}} \textbf{\bibinfo{volume}{91}}, \bibinfo{pages}{053001}
  (\bibinfo{year}{2003}).

\bibitem[{\citenamefont{Takamoto and Katori}(2003)}]{tka03}
\bibinfo{author}{\bibfnamefont{M.}~\bibnamefont{Takamoto}} \bibnamefont{and}
  \bibinfo{author}{\bibfnamefont{H.}~\bibnamefont{Katori}}, \bibinfo{journal}{{
  Phys. Rev. Lett.}} \textbf{\bibinfo{volume}{91}}, \bibinfo{pages}{223001}
  (\bibinfo{year}{2003}).

\bibitem[{\citenamefont{Derevianko et~al.}(2003)\citenamefont{Derevianko,
  Porsev, Kotochigova, Tiesinga, and Julienne}}]{dpk02}
\bibinfo{author}{\bibfnamefont{A.}~\bibnamefont{Derevianko}} \bibnamefont{{\it et al.}},
  \bibinfo{journal}{{ Phys. Rev. Lett.}} \textbf{\bibinfo{volume}{90}},
  \bibinfo{pages}{063002} (\bibinfo{year}{2003}).

\bibitem[{\citenamefont{Ido et~al.}(2000)\citenamefont{Ido, Isoya, and
  Katori}}]{iik00}
\bibinfo{author}{\bibfnamefont{T.}~\bibnamefont{Ido}},
  \bibinfo{author}{\bibfnamefont{Y.}~\bibnamefont{Isoya}}, \bibnamefont{and}
  \bibinfo{author}{\bibfnamefont{H.}~\bibnamefont{Katori}}, \bibinfo{journal}{{
  Phys. Rev. A}} \textbf{\bibinfo{volume}{61}}, \bibinfo{pages}{061403}
  (\bibinfo{year}{2000}).

\bibitem[{\citenamefont{Takusa et~al.}(2003)\citenamefont{Takusa, Maki, Komori,
  Takano, Honda, Kumakura, Yabuzaki, and Takahashi}}]{tmk03}
\bibinfo{author}{\bibfnamefont{Y.}~\bibnamefont{Takusa}} \bibnamefont{{\it et al.}},
  \bibinfo{journal}{{ Phys. Rev. Lett.}} \textbf{\bibinfo{volume}{91}},
  \bibinfo{pages}{040404} (\bibinfo{year}{2003}).

\bibitem[{\citenamefont{Ciurylo et~al.}()\citenamefont{Ciurylo, Tiesinga,
  Kotochigova, and Julienne}}]{ctk04}
\bibinfo{author}{\bibfnamefont{R.}~\bibnamefont{Ciurylo}},\bibnamefont{{\it et al.}},
 \bibinfo{note}{http://arXiv.org/physics/0407109}.

\bibitem[{\citenamefont{Derevianko}(2001)}]{der01}
\bibinfo{author}{\bibfnamefont{A.}~\bibnamefont{Derevianko}},
  \bibinfo{journal}{{ Phys. Rev. Lett.}} \textbf{\bibinfo{volume}{87}},
  \bibinfo{pages}{023002} (\bibinfo{year}{2001}).

\bibitem[{\citenamefont{Kokoouline et~al.}(2003)\citenamefont{Kokoouline,
  Santra, and Greene}}]{ksg03}
\bibinfo{author}{\bibfnamefont{V.}~\bibnamefont{Kokoouline}},
  \bibinfo{author}{\bibfnamefont{R.}~\bibnamefont{Santra}}, \bibnamefont{and}
  \bibinfo{author}{\bibfnamefont{C.~H.} \bibnamefont{Greene}},
  \bibinfo{journal}{{ Phys. Rev. Lett.}} \textbf{\bibinfo{volume}{90}},
  \bibinfo{pages}{253201} (\bibinfo{year}{2003}).

\bibitem[{\citenamefont{Santra and Greene}(2003)}]{sgr03}
\bibinfo{author}{\bibfnamefont{R.}~\bibnamefont{Santra}} \bibnamefont{and}
  \bibinfo{author}{\bibfnamefont{C.~H.} \bibnamefont{Greene}},
  \bibinfo{journal}{{ Phys. Rev. A}} \textbf{\bibinfo{volume}{67}},
  \bibinfo{pages}{062713} (\bibinfo{year}{2003}).

\bibitem[{\citenamefont{Degenhardt et~al.}(2003)\citenamefont{Degenhardt,
  Binnewies, Wilpers, Sterr, Riehle, Lisdat, and Tiemann}}]{dbw03}
\bibinfo{author}{\bibfnamefont{C.}~\bibnamefont{Degenhardt}} \bibnamefont{{\it et al.}},
  \bibinfo{journal}{{ Phys. Rev. A}} \textbf{\bibinfo{volume}{67}},
  \bibinfo{pages}{043408} (\bibinfo{year}{2003}).

\bibitem[{\citenamefont{Takusu et~al.}(2004)\citenamefont{Takusu, Komori,
  Honda, Kumakura, Yabuzaki, and Takahashi}}]{tkh04}
\bibinfo{author}{\bibfnamefont{Y.}~\bibnamefont{Takusu}} \bibnamefont{{\it et al.}},
  \bibinfo{journal}{{ Phys. Rev. Lett}} \textbf{\bibinfo{volume}{93}},
  \bibinfo{pages}{123202} (\bibinfo{year}{2004}).

\bibitem[{\citenamefont{Yasuda et~al.}(2004)\citenamefont{Yasuda, Kishimoto,
  Takamoto, and Katori}}]{ykt04}
\bibinfo{author}{\bibfnamefont{M.}~\bibnamefont{Yasuda}} \bibnamefont{{\it et al.}},
in
  \emph{\bibinfo{booktitle}{XIX International Conference on Atomic Physics, Rio
  de Janeiro}} (\bibinfo{year}{2004}).

\bibitem[{\citenamefont{Boutassetta et~al.}(1996)\citenamefont{Boutassetta,
  Allouche, and Aubert-Fr\'{e}con}}]{baa96}
\bibinfo{author}{\bibfnamefont{N.}~\bibnamefont{Boutassetta}},
  \bibinfo{author}{\bibfnamefont{A.~R.} \bibnamefont{Allouche}},
  \bibnamefont{and}
  \bibinfo{author}{\bibfnamefont{M.}~\bibnamefont{Aubert-Fr\'{e}con}},
  \bibinfo{journal}{{ Phys. Rev. A}} \textbf{\bibinfo{volume}{53}},
  \bibinfo{pages}{3845} (\bibinfo{year}{1996}).

\bibitem[{\citenamefont{Prodan et~al.}(2003)\citenamefont{Prodan, Pichler, Junker, Hulet and Bohn}}]{ppj03}
\bibinfo{author}{\bibfnamefont{I.} \bibnamefont{Prodan}} \bibnamefont{{\it et al.}},
  \bibinfo{journal}{{ Phys. Rev. Lett.}} \textbf{\bibinfo{volume}{91}},
  \bibinfo{pages}{080402} (\bibinfo{year}{2003}).

\bibitem[{\citenamefont{McKenzie et~al.}(2002)\citenamefont{McKenzie,
  Denschlag, Haffner, Browaeys, de~Araujo, Fatemi, Jones, Simsarian, Cho,
  Simoni et~al.}}]{mdh02}
\bibinfo{author}{\bibfnamefont{C.}~\bibnamefont{McKenzie}} \bibnamefont{{\it et al.}},
 \bibinfo{journal}{{ Phys. Rev. Lett.}}
  \textbf{\bibinfo{volume}{88}}, \bibinfo{pages}{120403}
  (\bibinfo{year}{2002}).

\bibitem[{\citenamefont{Katori et~al.}(1999)\citenamefont{Katori, Ido, Isoya,
  and Kuwata-Gonokami}}]{kii99}
\bibinfo{author}{\bibfnamefont{H.}~\bibnamefont{Katori}} \bibnamefont{{\it et al.}},
  \bibinfo{journal}{{ Phys. Rev. Lett.}} \textbf{\bibinfo{volume}{82}},
  \bibinfo{pages}{1116} (\bibinfo{year}{1999}).

\bibitem[{\citenamefont{Nagel et~al.}(2003)\citenamefont{Nagel, Simien, Laha,
  Gupta, Ashoka, and Killian}}]{nsl03}
\bibinfo{author}{\bibfnamefont{S.~B.} \bibnamefont{Nagel}} \bibnamefont{{\it et al.}},
 \bibinfo{journal}{{ Phys. Rev. A}}
  \textbf{\bibinfo{volume}{67}}, \bibinfo{pages}{011401}
  (\bibinfo{year}{2003}).

\bibitem[{\citenamefont{Fioretti et~al.}(1998)\citenamefont{Fioretti, Comparat,
  Crubellier, Dulieu, Masnou-Seeuws, and Pillet}}]{fcc98}
\bibinfo{author}{\bibfnamefont{A.}~\bibnamefont{Fioretti}} \bibnamefont{{\it et al.}},
  \bibinfo{journal}{{ Phys. Rev. Lett.}} \textbf{\bibinfo{volume}{80}},
  \bibinfo{pages}{4402} (\bibinfo{year}{1998}).

\bibitem[{\citenamefont{Kokoouline et~al.}(1999)\citenamefont{Kokoouline,
  Dulieu, Kosloff, and Masnou-Seeuws}}]{kdk99}
\bibinfo{author}{\bibfnamefont{V.}~\bibnamefont{Kokoouline}} \bibnamefont{{\it et al.}},
  \bibinfo{journal}{{J. Chem. Phys.}} \textbf{\bibinfo{volume}{110}},
  \bibinfo{pages}{9865} (\bibinfo{year}{1999}).

\bibitem[{\citenamefont{Kelly and Mathur}(1980)}]{kma80}
\bibinfo{author}{\bibfnamefont{F.~M.} \bibnamefont{Kelly}} \bibnamefont{and}
  \bibinfo{author}{\bibfnamefont{M.~S.} \bibnamefont{Mathur}},
  \bibinfo{journal}{{ Can. J. Phys.}} \textbf{\bibinfo{volume}{58}},
  \bibinfo{pages}{1416} (\bibinfo{year}{1980}).

\bibitem[{\citenamefont{Lurio et~al.}(1964)\citenamefont{Lurio, DeZafra, and
  Goshen}}]{ldg64}
\bibinfo{author}{\bibfnamefont{A.}~\bibnamefont{Lurio}},
  \bibinfo{author}{\bibfnamefont{R.~L.} \bibnamefont{DeZafra}},
  \bibnamefont{and} \bibinfo{author}{\bibfnamefont{R.~J.}
  \bibnamefont{Goshen}}, \bibinfo{journal}{{ Phys. Rev.}}
  \textbf{\bibinfo{volume}{134}}, \bibinfo{pages}{1198} (\bibinfo{year}{1964}).

\bibitem[{\citenamefont{Porsev et~al.}(2001)\citenamefont{Porsev, Kozlov,
  Rakhlina, and Derevianko}}]{pkr01}
\bibinfo{author}{\bibfnamefont{S.~G.} \bibnamefont{Porsev}} \bibnamefont{{\it et al.}},
  \bibinfo{journal}{{ Phys. Rev. A}} \textbf{\bibinfo{volume}{64}},
  \bibinfo{pages}{012508} (\bibinfo{year}{2001}).

\bibitem[{\citenamefont{Savukov and Johnson}(2002)}]{sjo02}
\bibinfo{author}{\bibfnamefont{I.~M.} \bibnamefont{Savukov}} \bibnamefont{and}
  \bibinfo{author}{\bibfnamefont{W.~R.} \bibnamefont{Johnson}},
  \bibinfo{journal}{{ Phys. Rev. A}} \textbf{\bibinfo{volume}{65}},
  \bibinfo{pages}{042503} (\bibinfo{year}{2002}).

\bibitem[{\citenamefont{Mitroy and Bromley}(2003)}]{mbr03}
\bibinfo{author}{\bibfnamefont{J.}~\bibnamefont{Mitroy}} \bibnamefont{and}
  \bibinfo{author}{\bibfnamefont{M.~W.~J.} \bibnamefont{Bromley}},
  \bibinfo{journal}{{ Phys. Rev. A}} \textbf{\bibinfo{volume}{68}},
  \bibinfo{pages}{052714} (\bibinfo{year}{2003}).

\bibitem[{\citenamefont{LeRoy}(1980)}]{ler80}
\bibinfo{author}{\bibfnamefont{R.~J.} \bibnamefont{LeRoy}}, in
  \emph{\bibinfo{booktitle}{Semiclassical Methods in Molecular Scattering and
  Spectroscopy}}, edited by \bibinfo{editor}{\bibfnamefont{M.~S.}
  \bibnamefont{Child}} (\bibinfo{publisher}{D. Reidel Publishing Company,
  Dordrecht, Holland}, \bibinfo{year}{1980}), pp. \bibinfo{pages}{109--126}.

\bibitem[{\citenamefont{Suominen et~al.}(1998)\citenamefont{Suominen, Band,
  Turvi, Burnett, and P.S.Julienne}}]{sbt98}
\bibinfo{author}{\bibfnamefont{K.-A.} \bibnamefont{Suominen}}
\bibnamefont{{\it et al.}}, \bibinfo{journal}{{Phys. Rev.
  A}} \textbf{\bibinfo{volume}{57}}, \bibinfo{pages}{3724}
  (\bibinfo{year}{1998}).

\bibitem[{\citenamefont{Loftus et~al.}(2004)\citenamefont{Loftus, Ido, Ludlow,
  Boyd, and Ye}}]{lil04}
\bibinfo{author}{\bibfnamefont{T.}~\bibnamefont{Loftus}} \bibnamefont{{\it et al.}},
\bibinfo{journal}{{
  Phys. Rev. Lett.}} \textbf{\bibinfo{volume}{93}}, \bibinfo{pages}{073003}
  (\bibinfo{year}{2004}).

\bibitem[{\citenamefont{C\^{o}t\'{e} et~al.}(2004)\citenamefont{C\^{o}t\'{e}, and Dalgarno}}]{cda98}
\bibinfo{author}{\bibfnamefont{R.}~\bibnamefont{C\^{o}t\'{e}}} \bibnamefont{and}
\bibinfo{author}{\bibfnamefont{A.} \bibnamefont{Dalgarno}},
\bibinfo{journal}{{
  Phys. Rev. A}} \textbf{\bibinfo{volume}{58}}, \bibinfo{pages}{498}
  (\bibinfo{year}{1998}).


\end{thebibliography}
\end{document}